\documentstyle[aps,preprint,prb]{revtex}
\begin{document}
\title{\bf A CONSTRAINT ON \\
THE ANOMALOUS GREEN'S FUNCTION}
\author{Yong-Jihn Kim$^{\dagger}$}
\address{Department of Physics, Purdue University, West Lafayette, Indiana 47907}
\maketitle
\def\hb{\hfill\break}
\begin{abstract}
It is shown that the physical constraint of the Anomalous Green's function
gives a natural pairing condition. 
The resulting self-consistency equation is directly related to the BCS gap 
equation. 
Both inhomogeneous and homogeneous systems are considered to illustrate
the importance of the constraint. 
Especially we find weak localization correction to the
phonon-mediated interaction.
\end{abstract}
\vskip 1pc
\noindent
$^{\dagger}$ Present address: Department of Physics, Korea Advanced 
Institute of Science 

and Technology, Taejon 305-701, Korea
\vfill\eject

\centerline{\bf 1. Introduction}
\vskip 1pc

Recently Green's function treatment of impurity effects on the superconductors
was shown to differ from that of the BCS theory.$^{1,2,3}$  
The electron density of states change caused by nonmagnetic impurities is overestimated
in the Abrikosov and Gor'kov's (AG) theory,$^{4}$ which is inconsistent
with Anderson's theorem.$^{5}$ (Details of the strong coupling theory of 
impure superconductors are discussed elsewhere.$^{6}$) 
For magnetic impurity effects, Kim and Overhauser$^{3}$ proposed a BCS type
theory with different predictions: (i) The initial slope of $T_{c}$
decrease depends on the superconductor and is not the universal
constant.$^{4}$ (ii) The $T_{c}$ reduction
by exchange scattering is partially suppressed by potential scattering
when the overall mean free path is smaller than the coherence length.
This compensation has been confirmed in several experiments.$^{7,8,9,10,11}$

In this letter we show that the physical constraint of the Anomalous
Green's function gives a natural pairing condition in Gor'kov's 
formalism.$^{12}$ 
The resulting self-consistency equation is nothing but another form of the 
BCS gap equation. Accordingly the above discrepancies are settled
in favor of the BCS theory.

To illustrate, we first consider inhomogeneous systems with nonmagnetic
and magnetic impurities. Weak localization correction to the
phonon-mediated interaction is derived.
We also discuss a homogeneous system.

\vskip 1pc
\centerline{\bf 2. Inhomogeneous System: Nonmagnetic Impurity Case}
\vskip 1pc

Let's consider an electron gas in the presence of the ordinary impurities. 
In the second-quantized representation, the Hamiltonian is
\begin{eqnarray}
H =  \int d{\bf{r}}\sum_{\alpha}\Psi^{\dagger}({\bf{r}}\alpha)[{\vec{p}^{2}\over 2m}
+ U_{o}({\bf{r}})]
\Psi({\bf{r}}\alpha) - {1\over 2}V\int d{\bf{r}}\sum_{\alpha\beta}\Psi^{\dagger}({\bf{r}}\alpha)
\Psi^{\dagger}({\bf{r}}\beta)\Psi({\bf{r}}\beta)\Psi({\bf{r}}\alpha). 
\end{eqnarray}
where $U_{o}({\bf{r}})=\sum_{i}U_{o}\delta({\bf{r}}-{\bf{R}}_{i})$ is the impurity potential.
The field operator, $\Psi({\bf{r}}\alpha)$, may
be expanded by the exact scattered states basis set $\psi_{n}({\bf{r}})$, that is,
\begin{eqnarray}
\Psi({\bf{r}}\alpha) = \sum_{n} \psi_{n}({\bf{r}}) c_{n\alpha},
\end{eqnarray}
where $c_{n\alpha}$ is a destruction operator of the electron.
The  Gor'kov's Anomalous Green's function $F({\bf{r}},{\bf{r}'})$ is 
defined$^{13}$
\begin{eqnarray}
F({\bf{r}}, {\bf{r}'}) = -<\Psi({\bf{r}}\uparrow)\Psi({\bf{r}'}\downarrow)>.
\end{eqnarray}
Accordingly, the pair potential is given
\begin{eqnarray}
\Delta({\bf{r}}) = VF({\bf{r}}, {\bf{r}}) = V\sum_{\omega}F({\bf{r}}, {\bf{r}}, \omega). 
\end{eqnarray}
The $\omega$'s are $\omega_{n} = (2n+1)\pi T$ for all integer $n$.
Near the transition temperature, it has been believed that$^{14}$ 
\begin{eqnarray}
F({\bf{r}}, {\bf{r}'}, \omega) = \int \Delta({\bf{l}})G^{\uparrow}_{\omega}({\bf{r}}, {\bf{l}})
G^{\downarrow}_{-\omega}({\bf{r}'}, {\bf{l}})d{\bf{l}},
\end{eqnarray}
where
\begin{eqnarray}
G^{\uparrow}_{\omega}({\bf{r},\bf{l}}) = \sum_{n}{\psi_{n\uparrow}({\bf{r}})\psi_{n\uparrow}^{*}({\bf{l}})
\over i\omega - \epsilon_{n}},  
\end{eqnarray}
and
\begin{eqnarray}
G^{\downarrow}_{-\omega}({\bf{r}',\bf{l}}) = \sum_{n'}{\psi_{n'\downarrow}({\bf{r}'})\psi_{n'\downarrow}^{*}({\bf{l}})
\over -i\omega - \epsilon_{n'}}.  
\end{eqnarray}
Substituting Eq. (5) in Eq. (4), we find a well-known self-consistency 
equation 
\begin{eqnarray}
\Delta({\bf{r}}) = VT\sum_{\omega}\int \Delta({\bf{l}})G^{\uparrow}_{\omega}({\bf{r},\bf{l}})
G^{\downarrow}_{-\omega}({\bf{r},\bf{l}})d\bf{l}.
\end{eqnarray}
\vskip 1pc
\centerline{\bf 2.1 A Constraint on the Anomalous Green's function}
\vskip 1pc

Now we consider the physical constraint of the Anomalous Green's function.
If we average over the impurity positions, the system becomes homogeneous.
Consequently, the Anomalous Green's function should be a function of 
${\bf{r}}-{\bf{r}'}$ after the impurity average, i.e.,
\begin{eqnarray}
\overline{F({\bf{r}}, {\bf{r}'})}^{imp} = \overline{F({\bf{r}}-{\bf{r}'})}^{imp}.
\end{eqnarray}
$\bar{\ }\bar{\ }^{imp}$ means an average over impurity positions
${\bf{R}}_{i}$.

However, if we substitute Eqs. (6) and (7) into Eq. (5) we find 
extra pairings between $n\uparrow$ and $n'(\not={\bar n})\downarrow$, which violate 
the physical constraint of the anomalous Green's function. 
$\bar{n}$ denotes the time reversed partner of the scattered state $n$.
Using the scattered states $\psi_{n(\vec{k})}(\bf{r})$,
\begin{eqnarray}
\psi_{n(\vec{k})}({\bf{r}}) = e^{i\vec{k}\cdot{\bf{r}}} + \sum_{\vec q}{U_{o}\over \epsilon_{\vec{k}} -
\epsilon_{\vec{k}+\vec {q}}}[\sum_{i}e^{-i\vec{q}\cdot{{\bf{R}}_{i}}}]e^{i(\vec{k}+\vec{q})\cdot {\bf{r}}}
+  \cdots ,
\end{eqnarray}
it is easy to show that
\begin{eqnarray}
\overline{\psi_{n(\vec{k})\uparrow}({\bf{r}})\psi_{n'(\vec{k}')\downarrow}({\bf{r}'})}^{imp} 
 &=& e^{i(\vec{k}\cdot{\bf{r}}+\vec{k}'\cdot{\bf{r} '})}[1 + V_{o}^{2}
\sum_{\vec{q},i}{e^{i\vec{q}\cdot({\bf{r}}-{\bf{r}'})}\over 
(\epsilon_{\vec{k}} - \epsilon_{\vec{k}+\vec{q}}) (\epsilon_{\vec{k}'} - \epsilon_{\vec{k}'-\vec{q}})}
+ \cdots ] \nonumber \\
 &\not=& f({\bf{r}}-{\bf{r}'}),
\end{eqnarray}
and
\begin{eqnarray}
\overline{\psi_{n(\vec{k})\uparrow}({\bf{r}})\psi_{{\bar n}(-\vec{k})\downarrow}({\bf{r}'})}^{imp} 
 &=& e^{i\vec{k}\cdot({\bf{r}}-{\bf{r} '})}[1 + V_{o}^{2}
\sum_{\vec{q},i}{e^{i\vec{q}\cdot({\bf{r}}-{\bf{r}'})}\over 
(\epsilon_{\vec{k}} - \epsilon_{\vec{k}+\vec{q}})^{2}} 
+ \cdots ]\nonumber \\
 &\equiv& g({\bf{r}}-{\bf{r}'}).
\end{eqnarray}
Note that even the zeroth order term of extra pairings between $n\uparrow$ and  
$n'(\not= {\bar n})\downarrow$ is not a function of ${\bf{r}}-{\bf{r}'}$.
From Eqs. (5) and (11) one finds that  
\begin{eqnarray}
\overline{F({\bf{r}},{\bf{r}'},\omega)}^{imp}&\sim& 
\overline{\psi_{n(\vec{k})\uparrow}({\bf{r}})\psi_{n'(\vec{k}')\downarrow}({\bf{r}'})}^{imp}
\nonumber\\
&\not=& \overline{F({\bf{r}}-{\bf{r}'},\omega)}^{imp}, 
\end{eqnarray}
in the presence of the extra pairing terms.

It is clear that only Anderson's pairing between $n\uparrow$ and ${\bar n}\downarrow$ is 
compatible with the homogeneity condition of the Anomalous Green's function
after averaging out the impurity positions.
Accordingly, the physical constraint of the Anomalous Green's function
gives a natural pairing condition. 
By eliminating the extra pairing, the `corrected' Anomalous Green's function
is given
\begin{eqnarray}
F({\bf{r}}, {\bf{r}'}, \omega) = \int \Delta({\bf{l}})\{G^{\uparrow}_{\omega}({\bf{r}}, {\bf{l}})
G^{\downarrow}_{-\omega}({\bf{r}'}, {\bf{l}})\}_{p.p.}d{\bf{l}},
\end{eqnarray}
where
\begin{eqnarray}
 \{G^{\uparrow}_{\omega}({\bf{r},\bf{l}})
G^{\downarrow}_{-\omega}({\bf{r}',\bf{l}})\}_{p.p.} = \sum_{n}
{\psi_{n\uparrow}({\bf{r}})\psi_{n\uparrow}^{*}({\bf{l}})\over i\omega - \epsilon_{n}}\sum_{n'}
{\psi_{n'\downarrow}({\bf{r}'})\psi_{n'\downarrow}^{*}({\bf{l}}) \over -i\omega - \epsilon_{n'}}
\delta_{n'={\bar n}},
\end{eqnarray}
and $p.p.$ means proper pairing constraint, which dictates pairing between
$n\uparrow$ and ${\bar n}\downarrow$. 
Consequently, the self-consistency equation also needs a proper pairing
constraint derived from the Anomalous Green's function.
The resulting equation is
\begin{eqnarray}
\Delta({\bf{r}}) = VT\sum_{\omega}\int \Delta({\bf{l}})\{G^{\uparrow}_{\omega}({\bf{r},\bf{l}})
G^{\downarrow}_{-\omega}({\bf{r},\bf{l}})\}_{p.p}d\bf{l}.
\end{eqnarray}

Let's compare Eq. (16) with the BCS gap equation,
\begin{eqnarray}
\Delta_{n} = -\sum_{n'}V_{nn'}{\Delta_{n'}\over 2\epsilon_{n'}}tanh({\epsilon_{n'}
\over 2T}),
\end{eqnarray}
where$^{3,15}$ 
\begin{eqnarray}
V_{nn'} = -V\int \psi^{*}_{n'}({\bf{r}})\psi^{*}_{\bar n'}({\bf{r}})
\psi_{\bar n}({\bf{r}})\psi_{n}({\bf{r}})d\bf{r}.
\end{eqnarray}
In fact, the relation between $\Delta({\bf{r}})$  
and $\Delta_{n}$ was obtained by Ma and Lee,$^{15}$ (i.e.), 
\begin{eqnarray}
\Delta_{n} = \int \psi_{n}^{*}({\bf {r}})\psi_{\bar n}^{*}({\bf {r}})
\Delta({\bf {r}})d\bf{r}.
\end{eqnarray}
Substitution of Eq. (16) into Eq. (19) leads to the BCS gap equation
Eq. (17), since$^{16}$
\begin{eqnarray}
T\sum_{\omega}{1\over \omega^{2} + \epsilon_{n}^{2}} = 
{1 \over 2\epsilon_{n}}tanh({\epsilon_{n}\over 2T}).
\end{eqnarray}

Note that because Eqs. (16) and (17) are basically equivalent,
they give the same result for the impurity effect on the superconductors.
Up to the first order in the impurity concentration, one finds$^{1,3}$ 
$V_{nn'}\cong -V$ and the density of states change is 
$O(1/E_{F}\tau)$. $E_{F}$ is the Fermi energy and $\tau$ is the scattering 
time. Therefore Anderson's theorem is obtained.
In the AG theory$^{4}$ the density of states change is 
$O(1/\omega_{D}\tau)$,$^{1}$ 
which gives a large decrease of $T_{c}$. $\omega_{D}$ is the Debye energy.
The Dyson expansion of Green's function in the presence of impurities
can not distinguish two energy scales $\omega_{D}$ and $E_{F}$ 
coming from the phonon-mediated interaction and the virtual impurity scattering. 

\vskip 1pc
\centerline{\bf 2.2 Weak localization correction}
\vskip 1pc
We are now ready to consider weak localization$^{17,18,19}$ correction
to the phonon-mediated interaction.
Mott and Kaveh$^{18}$ showed that the wavefunctions for the 
weakly localized states may be written as a mixture of power-law
and extended wavefunctions, of the form, for three
dimensions,
\begin{eqnarray}
\psi^{3d}_{n}({\bf{r}}) = A_{3}\Psi^{I}_{ext} + B_{3}{\Psi^{II}_{ext}\over
r^{2}},
\end{eqnarray}
and, for two dimensions,
\begin{eqnarray}
\psi^{2d}_{n}({\bf{r}}) = A_{2}\Psi_{ext}^{I} + B_{2}{\Psi_{ext}^{II}\over
r},
\end{eqnarray}
where
\begin{eqnarray}
A_{3}^{2}=1-4\pi B_{3}^{2}({1\over \ell}-{1\over L}), \quad B_{3} = {3\over 4\pi}{1\over k_{F}^{2}\ell},
\end{eqnarray}
and 
\begin{eqnarray}
A_{2}^{2}=1-2\pi B_{2}^{2}ln(L/\ell), \quad B_{2} = {1\over \pi^{2}k_{F}\ell}.
\end{eqnarray}
$\ell$ and $L$ are the elastic and inelastic mean free paths. 

We can find the effective interaction 
by substituting Eqs. (21) and (22) into Eq. (18). 
However, we must be careful in calculating  
contributions from the power-law wavefunction.
Because we are concerned with the bound state of
Cooper pairs in a BCS condensate, 
only the power-law wavefunctions within the 
the BCS coherence length $\xi_{o}$ are relevant.$^{3}$  
This is analogous to the insensitivity of the localized states with
the change of the boundary conditions.$^{20}$
Accordingly, it is obtained  
\begin{eqnarray}
V_{nn'}^{3d} &\cong&  -V[A_{3}^{4} + ({\xi_{o}^{3}\over \Omega})
8\pi A_{3}^{2}B_{3}^{2}({1\over \ell} -{1\over L}) + O(B^{4}_{3}\xi_{o}^{6}/
\Omega^{2})]\nonumber\\
&\cong& -V[1-{1\over (k_{F}\ell)^{2}}(1-{\ell\over L})],
\end{eqnarray}
\begin{eqnarray}
V_{nn'}^{2d} &\cong&  -V[A_{2}^{4} + ({\xi_{o}^{2}\over \Omega})
4\pi A_{2}^{2}B_{2}^{2}ln(\ell/L) + O(B^{4}_{2}\xi_{o}^{4}/\Omega^{2})]\nonumber\\
&\cong& -V[1-{2\over \pi k_{F}\ell}ln(L/\ell)].
\end{eqnarray}
Note that ${\xi_{o}^{3}/ \Omega}$ and $ {\xi_{o}^{2}/ \Omega}$
in the second terms of the right hand side of Eqs. (25) and (26)
denote the probabilities of finding the centers of the power-law wavefunctions
within the Cooper pair radius $\xi_{o}$ in 3 and 2 dimensions. 
$\Omega$ is the volume of the system.
It is remarkable that the same correction occurs in the conductivity and 
the phonon-mediated interaction by weak localization.
Consequently, one expects in 1 dimension
\begin{eqnarray}
V_{nn'}^{1d} \cong  
 -V[1-{1\over (\pi k_{F}a)^{2}}(L/\ell-1)],
\end{eqnarray}
where $a$ is the radius of the wire.

There are many experimental results which show the reduction of $T_{c}$
caused by weak localization.$^{21,22,23}$
Previously, it was interpreted by the enhanced Coulomb repulsion.$^{24}$
However, Dynes et al.$^{22}$ found a decrease of the Coulomb 
pseudo-potential $\mu^{*}$ with decreasing $T_{c}$.
We believe that this signals the importance of weak localization
correction to the phonon-mediated interaction.
Comparisons with the experiments will be published elsewhere.

\vskip 1pc
\centerline{\bf 3. Inhomogeneous System: Magnetic Impurity Case}
\vskip 1pc

We now consider the effect of magnetic impurities on the superconductivity.
Magnetic interaction is given by
\begin{eqnarray}
U_{m}({\bf{r}}) = \sum_{i}J\vec{s}\cdot\vec{S}_{i}\delta({\bf{r}} - {\bf{R}}_{i}), 
\end{eqnarray}
where $\vec{s} = {1\over 2}\vec \sigma$. 
The three components of $\vec \sigma$ are the Pauli matrices.
For convenience, we consider only the z-component of 
$U_{m}(\bf{r})$. The effect of each x and y-component is basically the same with 
that of z-component.$^{3}$
Notice that up and down spin electrons feel different potential because
of the Pauli spin matrix, $\sigma_{z}$, i.e.,
\begin{eqnarray}
[{\vec{p}^{2}\over 2m} + \sum_{i}{J\over 2}{\overline S}cos\theta 
\delta({\bf{r}} - {\bf{R}}_{i}) ]\psi_{n\uparrow}^{z}({\bf{r}}) = 
E_{n\uparrow}\psi_{n\uparrow}^{z}(\bf{r}),
\end{eqnarray}
\begin{eqnarray}
[{\vec{p}^{2}\over 2m} - \sum_{i}{J\over 2}{\overline S}cos\theta 
\delta({\bf{r}} - {\bf{R}}_{i}) ]\psi_{n\downarrow}^{z}({\bf{r}}) = 
E_{n\downarrow}\psi_{n\downarrow}^{z}(\bf{r}),
\end{eqnarray}
where ${\overline S} = \sqrt{S(S+1)}$ and $\theta$ is the polar angle of
the fixed local-spin $\vec S$.
As a result, one finds,$^{25}$
\begin{eqnarray}
\int {\psi_{n\uparrow}^{z \*}({\bf{r}})}^{*}\psi_{m\downarrow}^{z}({\bf{r}})d{\bf{r}} \not=
\delta_{nm}.
\end{eqnarray}

The Anomalous Green's function, near the transition temperature, is given
by$^{4}$
\begin{eqnarray}
F({\bf{r}}, {\bf{r}'}, \omega) = \int \Delta({\bf{l}})G^{\uparrow}_{\omega}({\bf{r}}, {\bf{l}})
G^{\downarrow}_{-\omega}({\bf{r}'}, {\bf{l}})d{\bf{l}},
\end{eqnarray}
where
\begin{eqnarray}
G^{\uparrow}_{\omega}({\bf{r},\bf{l}}) = \sum_{n}{\psi^{z}_{n\uparrow}({\bf{r}})
{\psi^{z}}^{*}_{n\uparrow}({\bf{l}})
\over i\omega - \epsilon_{n}},  
\end{eqnarray}
and
\begin{eqnarray}
G^{\downarrow}_{-\omega}({\bf{r}',\bf{l}}) = \sum_{n'}{\psi^{z}_{n'\downarrow}({\bf{r}'})
{\psi^{z}}^{*}_{n'\downarrow}({\bf{l}})
\over -i\omega - \epsilon_{n'}}.  
\end{eqnarray}
As in Sec. 2.1, we check the physical constraint of the Anomalous Green's
function 
\begin{eqnarray}
\overline{F({\bf{r}}, {\bf{r}'},\omega)}^{imp} = \overline{F({\bf{r}}-{\bf{r}'},\omega)}^{imp}.
\end{eqnarray}

Using the scattered states
\begin{eqnarray}
\psi_{n(\vec{k})\uparrow}^{z}({\bf{r}}) = e^{i\vec{k}\cdot{\bf{r}}} + \sum_{\vec{q}}{J{\overline S}
cos\theta/2\over \epsilon_{\vec{k}} -
\epsilon_{\vec{k}+\vec{q}}}[\sum_{i}e^{-i\vec{q}\cdot{\bf{R}}_{i}}]e^{i(\vec{k}+\vec{q})\cdot {\bf{r}}}
+  \cdots ,
\end{eqnarray}
and
\begin{eqnarray}
\psi_{n(\vec{k})\downarrow}^{z}({\bf{r}}) = e^{i\vec{k}\cdot{\bf{r}}} - \sum_{\vec{q}}{J{\overline S}
cos\theta/2\over \epsilon_{\vec{k}} -
\epsilon_{\vec{k}+\vec{q}}}[\sum_{i}e^{-i\vec{q}\cdot{\bf{R}}_{i}}]e^{i(\vec{k}+\vec{q})\cdot {\bf{r}}}
+  \cdots ,
\end{eqnarray}
it is straight forward to show that
\begin{eqnarray}
\overline{\psi^{z}_{n(\vec{k})\uparrow}({\bf{r}})\psi^{z}_{n'(\vec{k}')\downarrow}({\bf{r}'})}^{imp} 
 &=& e^{i(\vec{k}\cdot{\bf{r}}+\vec{k}'\cdot{\bf{r} '})}[1 - (J{\overline S}cos\theta/2)^{2}
\sum_{\vec{q},i}{e^{i\vec{q}\cdot({\bf{r}}-{\bf{r}'})}\over 
(\epsilon_{\vec{k}} - \epsilon_{\vec{k}+\vec{q}}) (\epsilon_{\vec{k}'} - \epsilon_{\vec{k}'-\vec{q}})}
+ \cdots ]\nonumber\\
 &\not=& f({\bf{r}}-{\bf{r}'}),
\end{eqnarray}
and
\begin{eqnarray}
\overline{\psi^{z}_{n(\vec{k})\uparrow}({\bf{r}})\psi^{z}_{{\bar n}(-\vec{k})\downarrow}({\bf{r}'})}^{imp} 
 &=& e^{i\vec{k}\cdot({\bf{r}}-{\bf{r} '})}[1 - (J{\overline S}cos\theta/2)^{2}
\sum_{\vec{q},i}{e^{i\vec{q}\cdot({\bf{r}}-{\bf{r}'})}\over 
(\epsilon_{\vec{k}} - \epsilon_{\vec{k}+\vec{q}})^{2}} 
+ \cdots ]\nonumber\\
 &\equiv& g({\bf{r}}-{\bf{r}'}).
\end{eqnarray}
The minus sign in the perturbation correction term of Eqs. (37) and (39) 
is the noteworthy feature of exchange scattering.
From Eqs. (35), (38) and (39), it is obvious that only the
pairing between the scattered states $n(\vec{k})\uparrow$ 
and ${\bar n}(-\vec{k})\downarrow$ is allowed.
Note that $n(\vec{k})\uparrow$ and ${\bar n}(-\vec{k})\downarrow$ are degenerate, 
\begin{eqnarray}
\overline {E_{n\uparrow}}^{imp} = \overline {E_{{\bar n}\downarrow}}^{imp}. 
\end{eqnarray}

The resulting self-consistency equation is 
\begin{eqnarray}
\Delta({\bf{r}}) = VT\sum_{\omega}\int \Delta({\bf{l}})\{G^{\uparrow}_{\omega}({\bf{r},\bf{l}})
G^{\downarrow}_{-\omega}({\bf{r},\bf{l}})\}_{p.p}d\bf{l}.
\end{eqnarray}
where
\begin{eqnarray}
 \{G^{\uparrow}_{\omega}({\bf{r},\bf{l}})
G^{\downarrow}_{-\omega}({\bf{r},\bf{l}})\}_{p.p.} = \sum_{n}
{\psi_{n\uparrow}^{z}({\bf{r}}){\psi^{z}}^{*}_{n\uparrow}({\bf{l}})\over i\omega - \epsilon_{n}}\sum_{n'}
{\psi^{z}_{n'\downarrow}({\bf{r}}){\psi^{z}}^{*}_{n'\downarrow}({\bf{l}}) \over -i\omega - \epsilon_{n'}}
\delta_{n'={\bar n}}.
\end{eqnarray}
Accordingly, we obtain the BCS gap equation 
\begin{eqnarray}
\Delta_{n} = -\sum_{n'}V_{nn'}{\Delta_{n'}\over 2\epsilon_{n'}}tanh({\epsilon_{n'}
\over 2T}),
\end{eqnarray}
where 
\begin{eqnarray}
V_{nn'} = -V\int {\psi^{z}}^{*}_{n'\uparrow}({\bf{r}})
{\psi^{z}}^{*}_{{\bar n'}\downarrow}({\bf{r}})
\psi^{z}_{{\bar n}\downarrow}({\bf{r}})\psi^{z}_{n\uparrow}({\bf{r}})d\bf{r}.
\end{eqnarray}

By taking an impurity average of Eq. (44), we find that
\begin{eqnarray}
\overline  {V_{nn'}}^{imp}
 = -V[1 - 2Z_{i}\sum_{\vec{q}}{(J{\overline S}cos\theta/2)^{2}\over 
(\epsilon_{\vec{k}} - \epsilon_{\vec{k}+\vec{q}})^{2}} 
  - 2Z_{i}\sum_{\vec{q}}{(J{\overline S}cos\theta/2)^{2}\over 
(\epsilon_{\vec{k}'} - \epsilon_{\vec{k}'+\vec{q}})^{2}}], 
\end{eqnarray}
where $Z_{i}$ is the number of the magnetic impurities.
Note that this result is the same as the z-component in Eq. (55) of 
Kim and Overhauser.$^{3}$
In other words, eliminating the extra pairing terms from the AG 
theory$^{4}$ gives rise to Kim and Overhauser's result.$^{3}$

\vskip 1pc
\centerline{\bf 4. Homogeneous System}
\vskip 1pc

Near the transition temperature, the Anomalous Green's function is given by
\begin{eqnarray}
F({\bf{r}}, {\bf{r}'}, \omega) = \int \Delta({\bf{l}})G^{\uparrow}_{\omega}({\bf{r}}, {\bf{l}})
G^{\downarrow}_{-\omega}({\bf{r}'}, {\bf{l}})d{\bf{l}},
\end{eqnarray}
where
\begin{eqnarray}
G^{\uparrow}_{\omega}({\bf{r},\bf{l}}) = \sum_{\vec{k}}{\phi_{\vec{k}\uparrow}({\bf{r}})\phi_{\vec{k}\uparrow}^{*}({\bf{l}})
\over i\omega - \epsilon_{\vec{k}}},  
\end{eqnarray}
and
\begin{eqnarray}
G^{\downarrow}_{-\omega}({\bf{r}',\bf{l}}) = \sum_{\vec{k}'}{\phi_{\vec{k}'\downarrow}({\bf{r}'})\phi_{\vec{k}'\downarrow}^{*}({\bf{l}})
\over -i\omega - \epsilon_{\vec{k}'}}.  
\end{eqnarray}
Note that $\phi_{\vec{k}}({\bf{r}}) = e^{i\vec{k}\cdot\bf{r}}$.
Accordingly, the self-consistency equation is
\begin{eqnarray}
\Delta({\bf{r}}) &=& VT\sum_{\omega}\int \Delta({\bf{l}})G^{\uparrow}_{\omega}({\bf{r},\bf{l}})
G^{\downarrow}_{-\omega}({\bf{r},\bf{l}})d\bf{l}\nonumber\\
&=&\int K({\bf{r}},{\bf{l}})\Delta({\bf{l}})d\bf{l}.
\end{eqnarray}

For a homogeneous system, Gor'kov$^{12}$ pointed out correctly that
$F({\bf{r}},{\bf{r}'})$ should depend only on ${\bf{r}}-{\bf{r}'}$, that is,
\begin{eqnarray}
F({\bf{r}},{\bf{r}'}) = F({\bf{r}}-{\bf{r}'}).
\end{eqnarray}
However, observe that
\begin{eqnarray}
\phi_{\vec{k}\uparrow}({\bf{r}})\phi_{\vec{k}'\downarrow}({\bf{r}'}) = e^{i(\vec{k}\cdot{\bf{r}} +
\vec{k}'\cdot{\bf{r}'})},
\end{eqnarray}
and
\begin{eqnarray}
\phi_{\vec{k}\uparrow}({\bf{r}})\phi_{-\vec{k}\downarrow}({\bf{r}'}) = e^{i\vec{k}\cdot({\bf{r}} -
{\bf{r}'})}.
\end{eqnarray}
Eqs. (46) and (49) include the extra pairing terms
between $\vec{k}\uparrow$ and $\vec{k}'\downarrow(\not= -\vec{k}\downarrow)$, which do not
satisfy the homogeneity condition of Eq. (50). In this case, the self-consistency
condition of the pair potential happens to eliminate the extra pairing
in Eq. (49) because of the orthogonality of the wavefunctions.
 
However, it is important to eliminate the extra pairing
in the Anomalous Green's function from the beginning.
Note that the kernel $K({\bf{r}},{\bf{l}})$ is not for the pairing 
between $\vec{k}\uparrow$ and $ -\vec{k}\downarrow$, but for the pairing between
the states which are the linear combination of the plane
wave states $\phi_{\vec{k}}({\bf{r}})$.$^{26}$ 
From the Bogoliubov-de Gennes equations, it can be shown that
the corresponding unitary transformation leads to the vacuum
state where the pairing occurs between the states, which are the
linear combination of the plane wave states.$^{26}$
The proper kernel for the BCS pairing between  
$\vec{k}\uparrow$ and $ -\vec{k}\downarrow$ is
\begin{eqnarray}
K^{c}({\bf{r}},{\bf{l}}) = VT\sum_{\omega} 
\{G^{\uparrow}_{\omega}({\bf{r}}, {\bf{l}}) G^{\downarrow}_{-\omega}({\bf{r}'}, {\bf{l}})\}_{p.p.},
\end{eqnarray}
where
\begin{eqnarray}
 \{G^{\uparrow}_{\omega}({\bf{r},\bf{l}})
G^{\downarrow}_{-\omega}({\bf{r}',\bf{l}})\}_{p.p.} = \sum_{\vec{k}}
{\phi_{\vec{k}\uparrow}({\bf{r}})\phi_{\vec{k}\uparrow}^{*}({\bf{l}})\over i\omega - \epsilon_{\vec{k}}}
\sum_{\vec{k}'}
{\phi_{\vec{k}'\downarrow}({\bf{r}'})\phi_{\vec{k}'\downarrow}^{*}({\bf{l}}) \over -i\omega - \epsilon_{\vec{k}'}}
\delta_{\vec{k}'={-\vec{k}}},
\end{eqnarray}
Consequently, the `corrected' Anomalous Green's function is
\begin{eqnarray}
F({\bf{r}}, {\bf{r}'}, \omega) = \int \Delta({\bf{l}})\{G^{\uparrow}_{\omega}({\bf{r}}, {\bf{l}})
G^{\downarrow}_{-\omega}({\bf{r}'}, {\bf{l}})\}_{p.p.}d{\bf{l}}.
\end{eqnarray}

In fact, Eq. (55) is not new.
Another form of this equation was already
obtained as shown$^{27, 28}$  
\begin{eqnarray}
F({\bf {r}}, {\bf {r}}') = \sum_{\vec{k}}{\Delta_{\vec{k}}\over 
2\epsilon_{\vec{k}}}tanh ({\epsilon_{\vec{k}}
\over 2T})exp{[i\vec{k}\cdot ({\bf{r}}-{\bf{r}}')]}.
\end{eqnarray}
By using 
\begin{eqnarray}
\Delta_{\vec{k}} = \int \phi_{\vec{k}}^{*}({\bf {r}})
\phi_{-\vec{k}}^{*}({\bf {r}})\Delta({\bf{r}})d\bf{r},
\end{eqnarray}
and
\begin{eqnarray}
T\sum_{\omega}{1\over \omega^{2} + \epsilon_{\vec{k}}^{2}} = 
{1 \over 2\epsilon_{\vec{k}}}tanh({\epsilon_{\vec{k}}\over 2T}),
\end{eqnarray}
Eq. (56) is transformed to 
\begin{eqnarray}
F(\bf{r}, \bf{r}') &=& \sum_{\omega}F(\bf{r}, \bf{r}', \omega),\nonumber\\ 
&=& \sum_{\omega} \int \Delta({\bf{l}})
\{G^{\uparrow}_{\omega}({\bf{r}}, {\bf{l}}) G^{\downarrow}_{-\omega}({\bf{r}'}, {\bf{l}})\}_{p.p.}d{\bf{l}}.
\end{eqnarray}
Therefore, Eq. (55) is confirmed.  

\vskip 1pc
\centerline{\bf 5. Conclusion}
\vskip 1pc
It is shown that the physical constraint of the Anomalous Green's function
gives a natural pairing constraint. The resulting self-consistency equation
is nothing but another form of the BCS gap equation.
Anderson's pairing between the
time reversed scattered state partners are obtained
in the presence of the ordinary impurities. 
Weak localization correction to the phonon-mediated interaction is calculated.
In the presence of the magnetic impurities, the degenerate partners are
paired, which give rise to the result of Kim and Overhauser. 
In the case of a homogeneous system, the BCS pairing is required.
\vskip 2pc
 
\centerline      {\bf ACKNOWLEDGMENTS} 

I am grateful to Professor A. W. Overhauser for discussions. 
The early version of this paper was circulated in the U.S.A., Japan, and
Korea from late 1994 to early 1995.
This work was supported by the National Science Foundation, Materials Theory
Program.
\vfill\eject
\centerline      {\bf REFERENCES} 
\vskip 1pt\hb
1. Yong-Jihn Kim and A. W. Overhauser, Phys. Rev. B{\bf 47}, 8025 (1993).\hb
2. Yong-Jihn Kim and A. W. Overhauser, Phys. Rev. B{\bf 49}, 12339 (1994).\hb
3. Yong-Jihn Kim and A. W. Overhauser, Phys. Rev. B{\bf 49}, 15779 (1994).\hb
4. A. A. Abrikosov and L. P. Gor'kov, Sov. Phys. JETP {\bf 12}, 1243 (1961).\hb
5. P. W. Anderson, J. Phys. Chem. Solids {\bf 11}, 26 (1959).\hb
6. Yong-Jihn Kim, ``Strong coupling theory of impure superconductors: 

real space formalism", to appear in Mod. Phys. Lett. B, 1996.\hb
7. M. F. Merriam, S. H. Liu, and D. P. Seraphim, Phys. Rev. {\bf 136}, A17 (1964).\hb
8. G. Boato, M. Bugo, and C. Rizzuto, Phys. Rev. {\bf 148}, 353 (1966).\hb
9. G. Boato and C. Rizzuto, (to be published), (referenced in A. J. Heeger, (1969), in

Solid State Physics {\bf 23}, eds. F. Seitz, D. Turnbull and H. Ehrenreich 
(Academic

 Press, New York), p. 409)  \hb
10. W. Bauriedl and G. Heim, Z. Phys. B{\bf 26}, 29 (1977); M. Hitzfeld and G. Heim, 

Sol. Sta.  Com. {\bf 29}, 93 (1979).\hb  
11. A. Hofmann, W. Bauriedl, and P. Ziemann, Z. Phys. B{\bf 46}, 117 (1982).\hb
12. L. P. Gor'kov, Sov. Phys. JETP {\bf 7}, 505 (1958); {\sl ibid.} {\bf 9}, 1364(1959).\hb
13. A. L. Fetter and J. D. Walecka, {\sl Quantum Theory of Many Particle Systems
} (McGraw-

Hill, New York, 1971), Section 31.\hb
14. A. A. Abrikosov, L. P. Gor'kov, and I. E. Dzyaloshinski, {\sl Methods of
Quantum Field 

Theory in Statistical Physics} (Prentice-Hall, Englewood, NJ, 1963), 
Eq. (38.3).\hb
15. M. Ma and P. A. Lee, Phys. Rev. B{\bf 32}, 5658 (1985).\hb
16. Murray R. Spiegel, {\sl Mathematical Handbook} (McGraw-Hill, New York, 
1968),

 formula 37.10.\hb
17. E. Abrahams, P. W. Anderson, D. C. Licciardello, and T. V. Ramakrishnan,
Phys. 

Rev. Lett. {\bf 42}, 673 (1979).\hb
18. N. F. Mott and M. Kaveh, Adv. Phys. {\bf 34}, 329 (1985).\hb
19. P. A. Lee and T. V. Ramakrishnan, Rev. Mod. Phys. {\bf 57}, 287 (1985).\hb
20. D. J. Thouless, Phys. Rep. {\bf 13C}, 93 (1974).\hb
21. G. Deutscher, A. Palevski, and R. Rosenbaum, {\sl Localization, Interaction,
and

Transport Phenomena}, ed. B. Kramer, G. Bergmann and Y. Bruynseraede 

(Springer, 1985), p. 108.\hb 
22. R. C. Dynes, A. E. White, J. M. Graybeal, and J. P. Garno, Phys. Rev. Lett.
{\bf 57}, 

2195 1986.\hb
23. D. B. Haviland, Y. Liu, T. Wang, and A. M. Goldman, Physica B {\bf 169}, 238 (1991),

and references therein.\hb
24. H. Fukuyama, Physica {\bf B}126, 306 (1984).\hb 
25. P. G. de Gennes and G. Sarma, J. Appl. Phys. {\bf 34}, 1380 (1963).\hb
26. Yong-Jihn Kim, unpublished.\hb
27. A. J. Leggett, Rev. Mod. Phys. {\bf 47}, 331 (1975), Eq. (6.45);
 C. Kittel, {\sl Quantum 

Theory of Solids} (John Wiley \& Sons, New York, 1963) Eq. (136) of ch. 8.\hb 
28. P. W. Anderson and P. Morel, Phys. Rev. {\bf 123}, 1911 (1961), Eq. (C.1).\hb 
\end{document}